# Transfer Learning across Different Chemical Domains: Virtual Screening of Organic Materials with Deep Learning Models Pretrained on Small Molecule and Chemical Reaction Data

Chengwei Zhang[1], Yushuang Zhai[1], Ziyang Gong[2], Hongliang Duan[3], Yuan-Bin She[1], Yun-Fang Yang[1], An Su[1, 2, *]

1. State Key Laboratory Breeding Base of Green Chemistry-Synthesis Technology, Key Laboratory of Green Chemistry-Synthesis Technology of Zhejiang Province, College of Chemical Engineering, Zhejiang University of Technology, Hangzhou, Zhejiang 310014, China

2. Key Laboratory of Pharmaceutical Engineering of Zhejiang Province, National Engineering Research Center for Process Development of Active Pharmaceutical Ingredients, Collaborative Innovation Center of Yangtze River Delta Region Green Pharmaceuticals, Zhejiang University of Technology, Hangzhou, 310014, P. R. China

3. Faculty of Applied Sciences, Macao Polytechnic University, Macao, 999078, China

**\*Corresponding Authors**

Prof. An Su

Associate Professor

College of Chemical Engineering

Zhejiang University of Technology

https://orcid.org/0000-0002-6544-3959

Email: ansu@zjut.edu.cn

# ABSTRACT

Machine learning is becoming a preferred method for the virtual screening of organic materials due to its cost-effectiveness over traditional computationally demanding techniques. However, the scarcity of labeled data for organic materials poses a significant challenge for training advanced machine learning models. This study showcases the potential of utilizing databases of drug-like small molecules and chemical reactions to pretrain the BERT model, enhancing its performance in the virtual screening of organic materials. By fine-tuning the BERT models with data from five virtual screening tasks, the version pretrained with the USPTO-SMILES dataset achieved $R^2$ scores exceeding 0.94 for three tasks and over 0.81 for two others. This performance surpasses that of models pretrained on the small molecule or organic materials databases and outperforms three traditional machine learning models trained directly on virtual screening data. The success of the USPTO-SMILES pretrained BERT model can be attributed to the diverse array of organic building blocks in the USPTO database, offering a broader exploration of the chemical space. The study further suggests that accessing a reaction database with a wider range of reactions than the USPTO could further enhance model performance. Overall, this research validates the feasibility of applying transfer learning across different chemical domains for the efficient virtual screening of organic materials.

## KEYWORDS

# INTRODUCTION

Organic materials, such as organic photovoltaics (OPVs), organic light-emitting diodes (OLEDs), and organic redox flow batteries (ORFBs), play a crucial role in materials science and their unique functional properties are widely utilized in various research fields[1-4]. Since organic materials have complex structures, it is often costly to explore them through wet experiments. Therefore, virtual screening helps to screen target organic materials before conducting wet experiments. Starting from the simplest Quantitative Structure-Activity Relationships (QSAR) models, virtual screening has a long history in the field of drug discovery.[5-8] Nowadays, in the era of artificial intelligence, different types of state-of-the-art (SOTA) machine learning model architectures have been developed for the virtual screening of drug-like small molecules, including Transformers[9], graph neural networks[10,11], sequence-graph hybrid models[12], and large language models (LLMs)[13]. On the other hand, for organic materials, the Aspuru-Guzik group proposed the concept of a "computational funnel" for the virtual screening of organic materials in 2015, suggesting that machine learning can be used as an inexpensive and efficient layer of virtual screening that can be placed before more expensive methods such as Density Functional Theory (DFT) calculation.[14] While there have been several studies utilizing machine learning methods for virtual screening of organic materials[15-18], they could not avoid one major challenge - the limited amount of training data.

There are several ways to address the problem of scarce training data. Some researchers have chosen to augment the dataset by supplying a wider range of physical and chemical information at the molecular level.[19-21] Another common strategy is transfer learning[18,22,23], a popular machine learning technique that involves pre-training with a large dataset and fine-tuning using smaller datasets, which allows for efficient learning using limited resources. However, traditional transfer learning based on supervised learning requires similar types of targets in the pre-training dataset and fine-tuning datasets. Otherwise, the transfer learning may produce results opposite to those expected.[24] On the other hand, recent studies suggest that the Bidirectional Encoder Representations from Transformers (BERT) model[25] with an unsupervised pre-training phase and a supervised fine-tuning phase may be able to address the limitation of supervised transfer learning in chemistry. Schwaller et al. developed and successfully used a BERT-based framework, *rxnfp*, for predictive chemical reaction classification, which pre-trained and fine-tuned the BERT models using databases consisting of different classification systems.[26] In our group's previous studies on PorphyBERT [17] and SolvBERT[27], we found that unsupervised learning using BERT models and their subsequent fine-tuning allows the model to learn chemistry from a large number of molecules without being affected by differences in the targets of the pre-training dataset and fine-tuning datasets.

Since the pre-training phase of the BERT model involves unsupervised learning and does not require any property or activity data of the molecules, it is theoretically possible to incorporate a large number of molecular structures into the pre-training process, thereby generating a pre-trained model with a wealth of knowledge in a chemical space larger than one with only organic materials. Therefore, in addition to the organic materials database, drug-like small molecule data can be included in the pre-training data, which could potentially broaden the model's horizon for the variety of organic parts of the organic materials. Meanwhile, chemical reaction databases can offer a wider range of chemical structures, including organic and inorganic materials, metals, complexes, and molecular associations. To the best of our knowledge, few studies are exploring whether chemical reaction data can be used as pretraining data for deep learning models to predict molecular properties.

In this study, we first pretrained the BERT models using large databases such as chemical reaction data, drug-like small molecule data, and organic materials data. After pretraining, we fine-tuned these models using small organic materials databases for different prediction tasks. In addition, to further explain the performance of the models trained on different combinations of the pretraining and fine-tuning databases, we summarized their statistics of organic build blocks and visualized their chemical space. Furthermore, we investigated the effect of the size of the pretraining and fine-tuning datasets on the predictive performance of the models. Finally, we compared the best models proposed in this study with the models without an unsupervised pretraining phase and with the Transformer/BERT models pretrained using other databases.

## METHODS

### Datasets for pretraining

**ChEMBL.** ChEMBL is a manually curated database of bioactive molecules with drug-like properties. It brings together chemical, bioactivity, and genomic data to help translate genomic information into effective new drugs.[28] The data used in this study were obtained from the ChEMBL download channel (https://www.ebi.ac.uk/chembl, accessed April 10, 2023), which contains Simplified Molecular Input Line Entry System (SMILES)[29] data for 2,327,928 drug-like small molecules.

**Clean Energy Project Database (CEPDB).** In 2008, Aspuru-Guzik et al. launched the Harvard Clean Energy Project (CEP) to help find high-efficiency organic photovoltaic materials.[30] They built the main CEP using a combinatorial molecular generator, and by 2019, the project team had synthesized at least 2,322,849 molecules. In this study, CEPDB data was downloaded from https://figshare.com/articles/dataset/moldata_csv/9640427 (accessed April 12, 2023), from which $10^4$, $10^5$, and $10^6$ molecules were randomly selected as training datasets, referred to as CEPDB-10K, CEPDB-100K, and CEPDB-1M, respectively.

**United States Patent and Trademark Office (USPTO) databases.** The USPTO

database

([https://figshare.com/articles/dataset/Chemical_reactions_from_US_patents_1976-Sep2016_/5104873](https://figshare.com/articles/dataset/Chemical_reactions_from_US_patents_1976-Sep2016_/5104873), accessed on April 13, 2023) contains reactions extracted through text-mining from U.S. patents published between 1976 and September 2016, available as reaction SMILES. The USPTO database used in this study was derived from the study by Giorgio Pesciullesi et al.[22] and contains 1,048,575 reactions. In addition, the molecules were extracted from these chemical reactions, resulting in 5,390,894 molecules that comprised the USPTO-SMILES dataset. Furthermore, duplicate molecules were removed from the molecules in the USPTO-SMILES dataset to obtain the USPTO-SMILES-clean dataset, which contains 1,345,854 molecules with different SMILES.

**Datasets for fine-tuning and evaluation**

**Metalloporphyrin and porphyrin Database (MpDB).** Porphyrins are a class of heterocyclic macrocycle organic compounds containing four modified pyrrole substituents interconnected by methyl bridges (=CH−) at their α-carbon atoms. Porphyrins are the active central structure of chlorophyll and have tunable photochemical properties when coordinated with metal ions. MpDB is derived from the Computational Materials Repository database (CMR) of porphyrin-based dyes database[31,32] ([https://cmr.fysik.dtu.dk/dssc/dssc.html](https://cmr.fysik.dtu.dk/dssc/dssc.html), accessed on March 23, 2023) and contains for 12,096 porphyrins or metalloporphyrin with structural and energy level information. The porphyrin dye structures in MpDB were converted to canonical SMILES using our previously developed framework[17]. The HOMO-LUMO gap was used as a virtual screening task for the models fine-tuned with MpDB.

**Phthalocyanine Database (PcDB).** Phthalocyanines are large, aromatic, macrocyclic, and fully synthetic organic compound with structures similar to porphyrins, which are also of theoretical or specialized interest in chemical dyes and photovoltaics. In this study, we curated the PcDB by searching from the PubChem[33] database using the

keyword "Phthalocyanine". The total number of molecules in this database was 848. The virtual screening task was XlogP (computed octanol-water partition coefficient) and there were only 481 molecules with this property. Given the constrained size of the dataset, we applied a 5-fold SMILES data augmentation[34] to the PcDB dataset following its segmentation into training, evaluation, and test subsets.

**Experimental database of Optical properties of Organic compounds (EOO).** Organic chromophores are an important class of compounds in photochemistry. In 2020, Joung et al. curated a database of 20,236 data points on the optical properties of organic compounds from literature[35] containing 7,016 unique organic chromophores in 365 solvents or solids. In this study, max absorption wavelength (MAW) and max emission wavelength (MEW) were two virtual screening tasks with 17,295 and 18,142 data points, respectively.

**Database of Organic donor-acceptor molecules (solar).** Organic donor-acceptor molecules are an important class of compounds in the field of organic photochemistry due to their unique electronic and optical properties. A database of organic donor-acceptor molecules is publicly available in the CMR (https://cmr.fysik.dtu.dk/solar/solar.html, accessed on March 23, 2023), which contains the Kohn-Sham (B3LYP) HOMO, LUMO, HOMO-LUMO gap, and singlet-triplet gap of organic donor-acceptor molecules. We selected Kohn-Sham HOMO-LUMO gap as the label for the virtual screening task, with a total of 5,366 data.

**Models**

**BERT.** BERT (Bidirectional Encoder Representations from Transformers) is a pretrained natural language processing model developed by Google[25]. BERT has two distinct phases: pretraining and fine-tuning. During pretraining, BERT learns about the contextual relationships between words and sentences through unsupervised learning on large amounts of text data. Fine-tuning involves additional training by virtual screening classification or regression tasks. In this way, the pretrained model is better able to predict the specific nuances of the virtual screening task, ultimately achieving higher performance. Fine-tuning BERT typically requires the use of only a small amount of task-specific data and is therefore much faster than pretraining.

The BERT model architecture used in this study was built based on the *rxnfp* framework[26]. The pretraining data was randomly divided into two parts: training set and validation set, with a ratio of 9:1 and training epochs of 20. During pretraining, only the SMILES of molecules were provided to the model, and the model learned the structural information of molecules in an unsupervised learning way. For fine-tuning, the regression task dataset used for the fine-tuning phase was randomly split into training, validation, and test sets in a fine-tuning ratio of 8:1:1. Each fine-tuning was done with 50 training epochs and the learning rate was set to $10^{-5}$. The most proficient pretrained model underwent further optimization in the fine-tuning stage, with specific adjustments made to the learning rate and the dropout probability of neurons within the hidden layers.

**Baseline models.** Two classical machine learning models, Random Forest (RF) and Support Vector Machine Regression (SVR) were used as baseline models. The molecular representation chosen for these two models was MAP4, a molecular fingerprint recently developed by Reymond's group[36]. In addition, a graph convolutional network (GCN) developed by Shen et al.[37] was chosen as another baseline model. GCN reads the SMILES of a molecule and converts the molecule into a graph format. Each of the three baseline models was trained directly by the regression tasks on organic materials in a supervised learning way. Furthermore, DeepChem-77M is a pretrained model based on the RoBERTa[38] transformer using 77M unique SMILES from ChemBERTa[39] for pretraining. In this study, the model was directly fine-tuned by the regression tasks on organic materials.

**Chemical Space Visualization.**

The SMILES of the molecule are partially encoded by the unsupervised learning model during the pre-training process, generating a molecular fingerprint. The dimensionality of the fingerprint was reduced by TMAP, which reduced the high-dimensional data to a tree structure that can display local association and global distribution.[40] Finally, the Molecular fingerprints with reduced dimensions was visualized by Faerun[41].

# RESULTS AND DISCUSSIONS

In this study, we adopted BERT, one of the SOTA models in natural language processing, as the basic model architecture. The training of BERT was divided into a pretraining and a fine-tuning phase, as shown in Figure 1. The BERT models were pretrained using databases in different chemical domains, such as USPTO (chemical reactions), ChEMBL (drug-like small molecules), and CEPDB (organic materials), respectively. The pretrained models were then fine-tuned for different regression tasks in organic materials, including the band gap in MpDB, xlogP in PcDB, KS_gap in Solar, as well as the maximum absorption wavelength (MAW) and maximum emission wavelength (MEW) in EOO. It is worth noting that these five virtual screening tasks differ significantly in terms of the range of values and type of properties. The major hyperparameters for the pretraining and fine-tuning phases are listed in **Table S1** and **Table S2**.

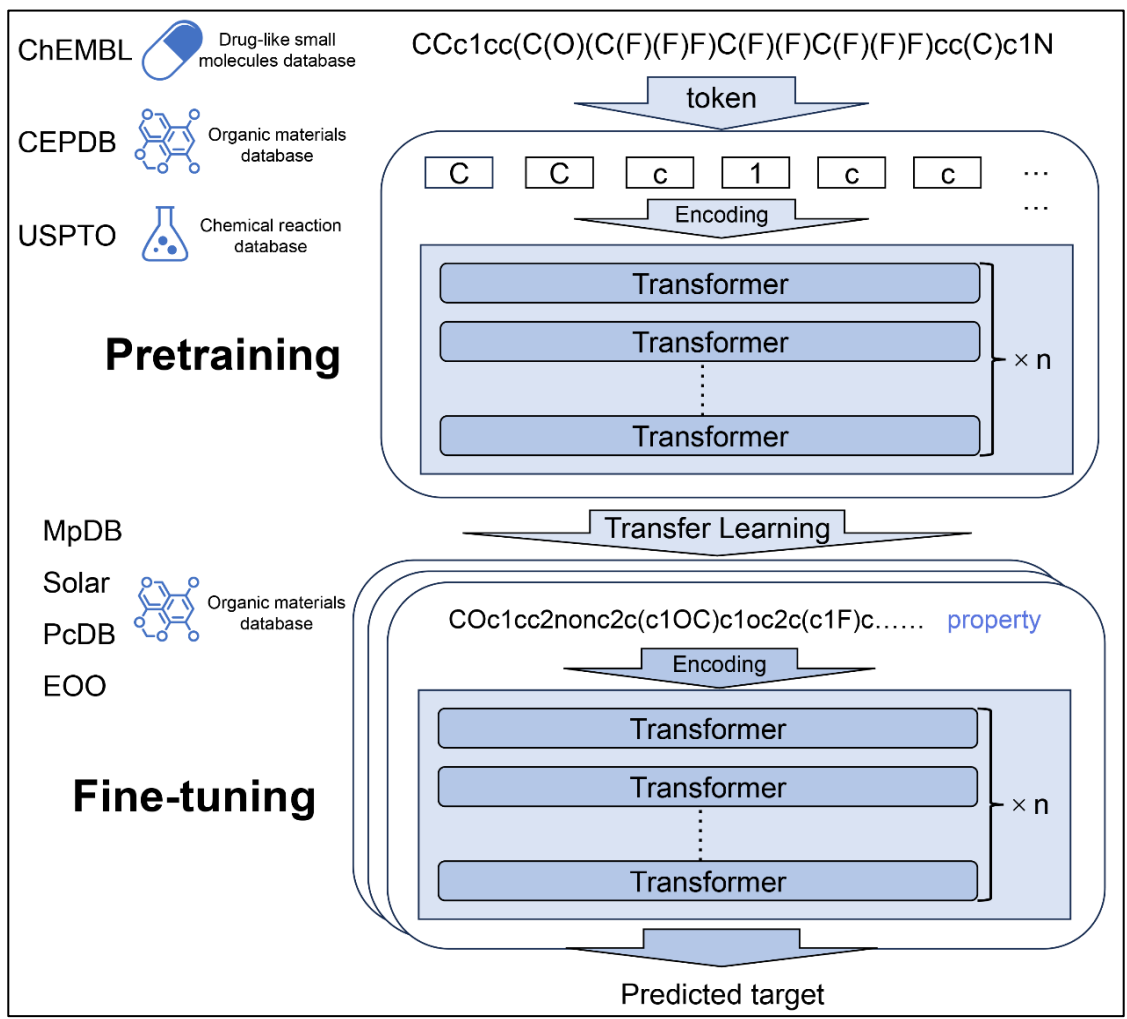

**Figure 1.** The workflow of this study. The models were pretrained using chemical databases from different chemical domains and then fine-tuned individually for different virtual screening tasks in organic materials.

To eliminate the possible effects of differences in the structure of Reaction SMILES (the original format of USPTO) and normal SMILES data, we further prepared two additional USPTO datasets by extracting the SMILES of all the reaction components from the USPTO dataset, generating USPTO-SMILES (Figure 2, step 1), and then further removing the duplicate molecular SMILES in USPTO-SMILES to generate USPTO-SMILES-clean (Figure 2, step 2).

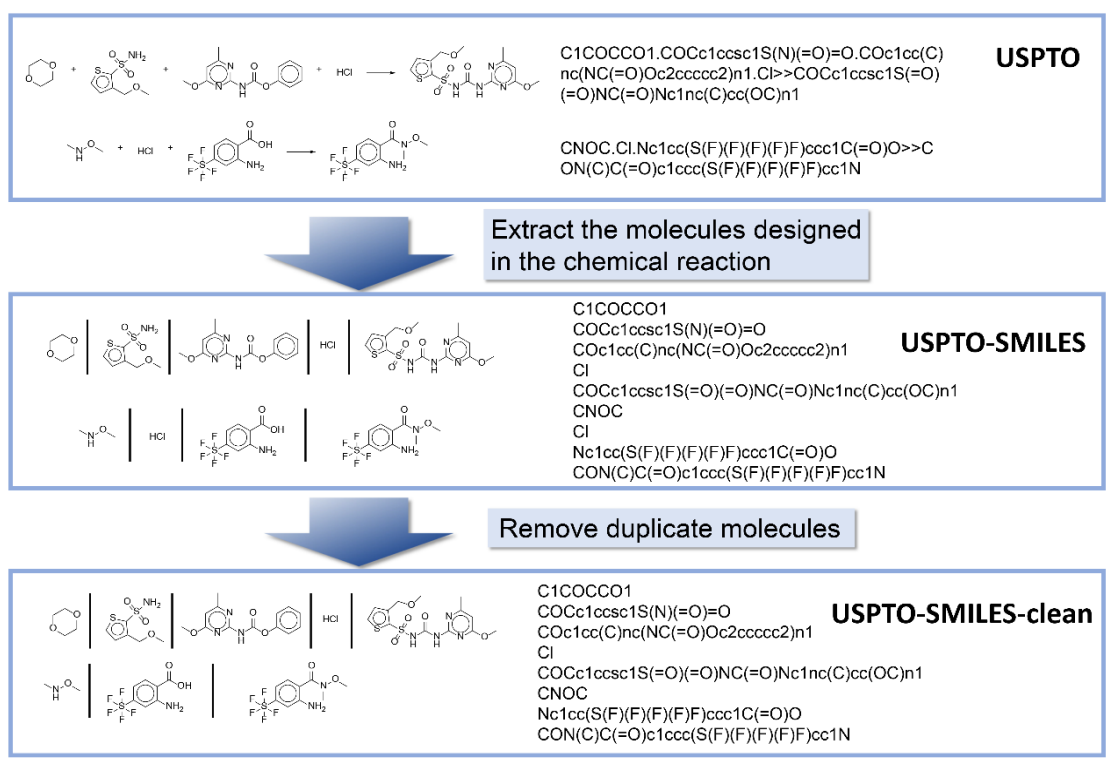

**Figure 2.** The workflow for creating the USPTO-SMILES and USPTO-SMILES-clean databases. The SMILES of molecules in USPTO were extracted to create USPTO-SMILES, while the duplicate SMILES in USPTO-SMILES were removed to create USPTO-SMILES-clean.

**Performance of pretrained BERT models for virtual screening tasks.** Figure 3 illustrates the prediction performance of the five models pretrained using different databases and subsequently fine-tuned with virtual screening tasks. Given the challenge of directly comparing absolute errors across tasks due to variations in their units, the evaluation primarily utilizes the $R^2$ metric, with detailed results on Mean Absolute Error (MAE) and Root Mean Square Error (RMSE) for each model provided in Tables S3 to S5. Among the evaluated models, those pretrained on the USPTO-SMILES database consistently achieved the highest performance across all five virtual screening tasks. Intriguingly, the model pretrained using the CEPDB database, which is the sole database dedicated to organic materials, exhibited the lowest performance in four out of the five tasks. This outcome underscores the significant impact of the choice of the pretraining database on the model's effectiveness in virtual screening applications.

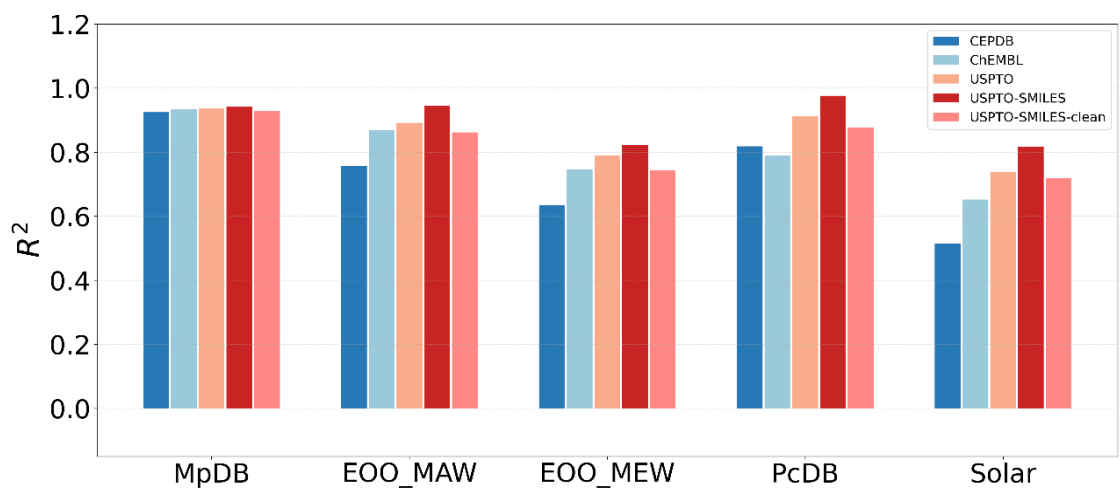

**Figure 3.** $R^2$ values of transfer learning using different databases for pretraining

Figure 3 illustrates that among the three models trained using differently processed USPTO data, those pretrained on USPTO-SMILES consistently outperform the model using raw USPTO data. This improvement highlights the benefits of extracting molecular SMILES from the raw data, which likely enhances pretraining by eliminating non-structural information (Figure 2). Moreover, the USPTO-SMILES pretrained model also surpasses the performance of the model pretrained on USPTO-SMILES-clean, despite both datasets sharing identical data structures and covering similar chemical spaces. This superiority can be attributed to the USPTO-SMILES dataset being four times larger than its clean counterpart. The repetition of SMILES in the USPTO-SMILES dataset often represents molecules that are more commonly encountered in chemical reactions. This repetition aids in a deeper understanding of chemical language, suggesting that a larger, more repetitive dataset can be beneficial for model training in this context.

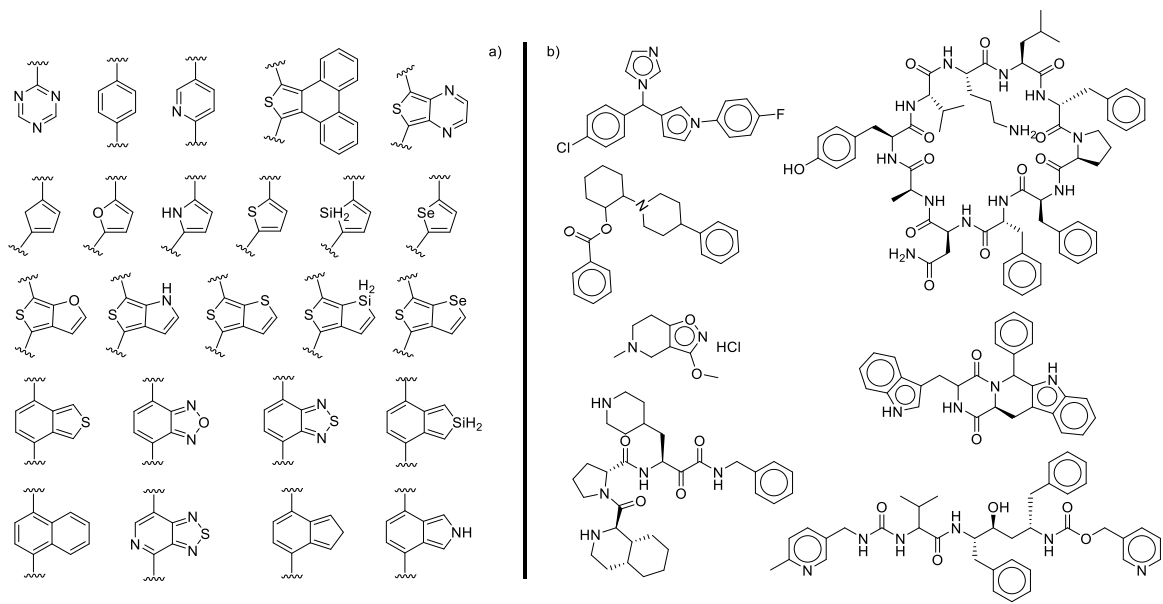

**Figure 4.** a) The 26 building blocks used for generating the CEPDB; b) Molecules in the ChEMBL database are displayed, selected at random from the database.

**Statistics of organic building blocks in the databases.** The superior performance of models pretrained with the USPTO-SMILES and ChEMBL databases over those using the CEPDB, an organic material database, can be attributed to the richer diversity of organic building blocks available in the former two (Figure 4). Utilizing RDKit's Chem.Fragments package, we quantified the number of common organic building blocks across the pretraining and fine-tuning databases, with these findings detailed in Figure 5 and further elaborated in Table S6. Figures 5a to 5c reveal that USPTO-SMILES and ChEMBL contain a significantly broader spectrum of organic building blocks compared to CEPDB. Specifically, of the 85 identified organic building blocks, CEPDB lacked 72, whereas USPTO-SMILES and ChEMBL were missing only one. This extensive repository of organic building blocks in USPTO-SMILES and ChEMBL is crucial for the enhanced performance of the BERT models pretrained on these databases for the virtual screening tasks.

**Figure 5.** Statistics of organic building blocks in a) CEPDB, b) ChEMBL, c) USPTO-SMILES, d) MpDB, e) EOO_MAW, and f) Solar. For teh details of these organic building blocks, an alternative tabular form of this summary is shown in Table S6.

**Chemical Space Coverage.** We aimed to explore the extent of chemical space encompassed by the databases used for pretraining and fine-tuning our models. To achieve this, we applied TMAP[40] for dimensionality reduction on the fingerprints generated by the USPTO-SMILES model and visualized the resultant chemical space using Faerun[41]. As depicted in Figure 6, the chemical space covered by USPTO-SMILES (highlighted in red) is notably broader than that covered by both ChEMBL (in green) and CEPDB (in blue). This broader coverage by USPTO-SMILES is more comprehensively explored through the interactive TMAP visualizations available in the S.I. The expansive chemical space coverage of USPTO-SMILES accounts for its pretrained models outperforming those pretrained on ChEMBL, despite both databases featuring a similar diversity of organic building blocks. This superior performance can be attributed to USPTO-SMILES not only encompassing drug-like small molecules found in ChEMBL but also including metals, non-metallic inorganic compounds, and organic materials frequently used as catalysts or reagents in reactions.

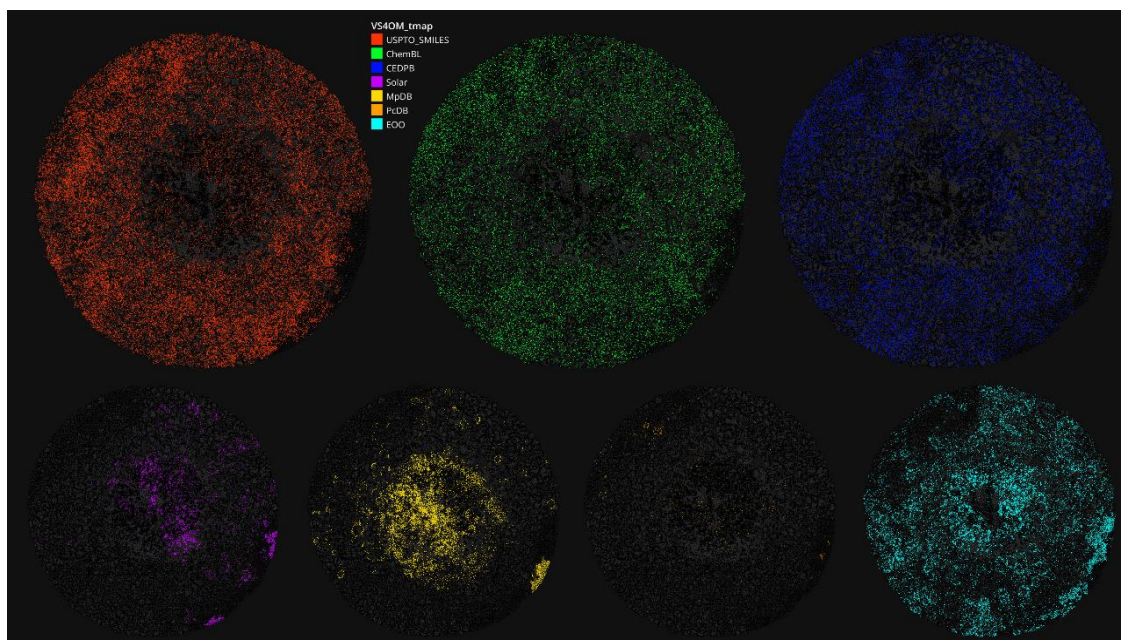

**Figure 6.** The two-dimensional chemical space of USPTO-SMILES, CEPDB, ChEMBL, Solar,

MpDB, PcDB, and EOO. The molecules representation, dimensionality reduction, and visualization were implemented by BERT, TMAP[40], and Faerun[41], respectively. To maximize the visualization, 50,000, 20,000, and 20,000 data were randomly selected from USPTO-SMILES, ChEMBL, and CEPDB, respectively, for visualization.

**Effect of the relative size of pretraining and fine-tuning datasets.** To better comprehend the impact of pretraining and fine-tuning dataset sizes on prediction performance, we randomly selected 2 million to 10,000 data instances from USPTO-SMILES and 1 million to 10,000 data instances from ChEMBL and CEPDB, respectively, and pretrained the models using these sampled datasets (Figure ). Within the framework of the USPTO-SMILES and ChEMBL pre-training datasets, the $R^2$ for four out of the five tasks assessed typically demonstrated a decreasing trend with the reduction in the size of the pre-training dataset. Conversely, this decline was not evident in the tasks pre-trained on the CEPDB dataset, likely a consequence of the previously discussed insufficiency in chemical information inherent to CEPDB.

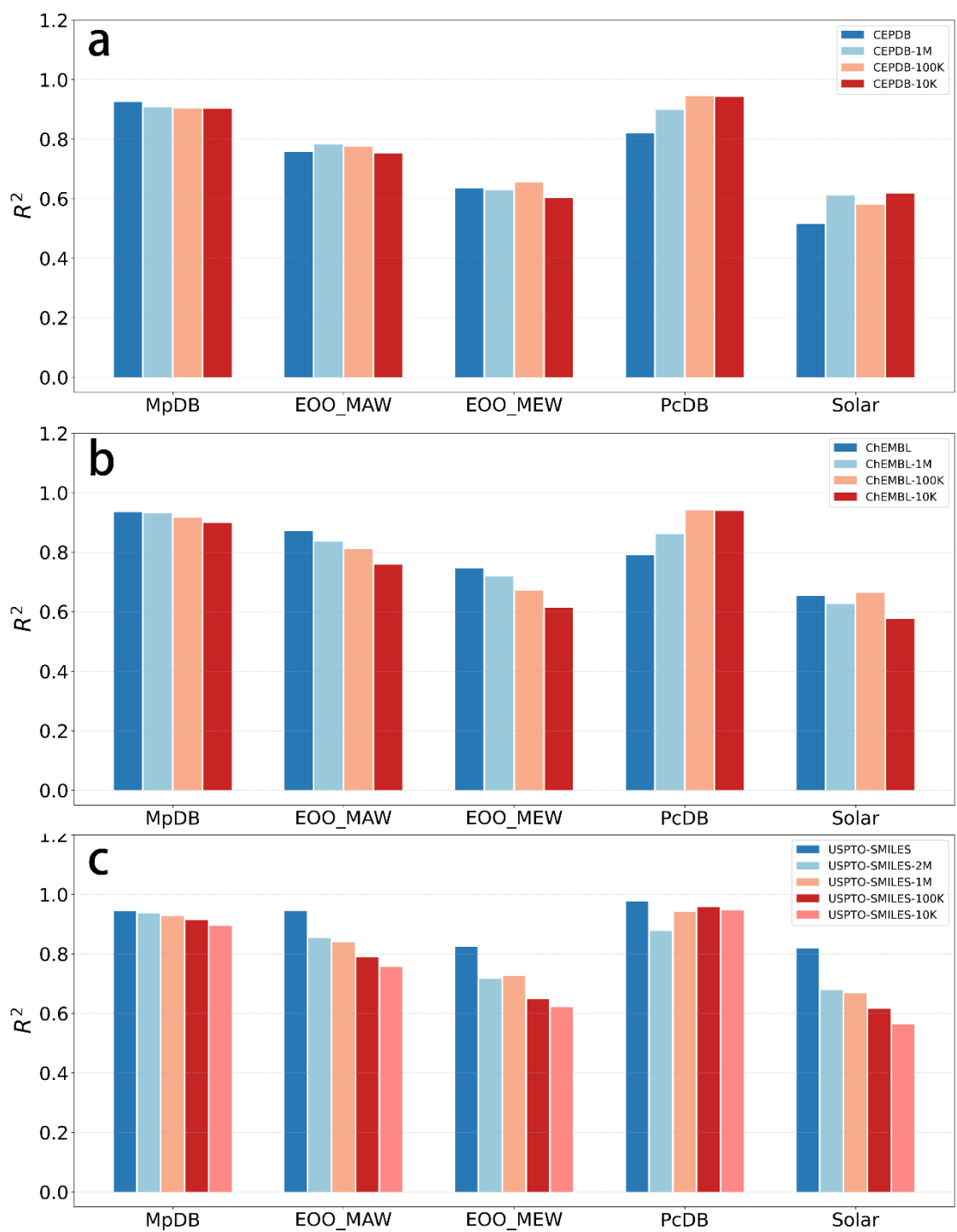

**Figure 7.** $R^2$ value of five virtual screening prediction tasks with models pretrained using random samples of (a) CEPDB, (b) ChEMBL, and (c) USPTO-SMILES of different sizes

**Comparison with the baseline models.** To further understand the predictive capability of USPTO-pretrained models for the virtual screening of organic materials, the best-performed USPTO-pretrained models in this study were compared with four baseline models, encompassing two traditional machine learning models, a graph convolution neural network (GCN)[37], and the DeepChem-77M[44,39] model trained on a Roberta architecture (**Table 1**). The findings indicate that our USPTO-SMILES model surpasses all competitors across four of the five evaluated tasks, underscoring the superiority of our approach in virtual screening applications. The sole deviation was observed with the PcDB dataset, where the GCN model achieved a marginally higher ($R^2$) by 0.02, likely attributable to the dataset's limited size, which potentially aligns more closely with the GCN's architectural strengths[42]. Furthermore, the comparative underperformance of DeepChem-77M could be ascribed to its pre-training on the PubChem database, which offers a narrower chemical space than USPTO as discussed above, and its adoption of a Roberta architecture with fewer hidden layers than BERT, potentially diminishing its performance.

**Table 1.** The performance comparison between the USPTO-SMILES model and the baseline models

| $R^2$ | MpDB | EOO_MAW | EOO_MEW | PcDB | Solar |
|---|---|---|---|---|---|
| USPTO-SMILES (*our study*) | **0.9428** | **0.9443** | **0.8229** | 0.9759 | **0.8174** |
| Support Vector Regressor (SVR) | 0.9320 | 0.2086 | 0.2187 | 0.4149 | 0.7714 |
| Random Forest (RF) | 0.8953 | 0.8959 | 0.8081 | 0.9981 | 0.8416 |
| Graph Convolution Neural Network[43] (GCN) | 0.8950 | 0.8384 | 0.6809 | **0.9983** | 0.8045 |
| DeepChem-77M[44] | 0.9097 | 0.8098 | 0.6457 | 0.9571 | 0.6158 |

Our approach to enhancing virtual screening methodologies could see significant improvements by incorporating a commercial database like Pistachio, which offers a more comprehensive collection of reaction data than the USPTO by including reactions from the European Patent Office (EPO). Access to Pistachio directly is, unfortunately, not possible for our team. However, we utilized the open-source rxnfp-Pistachio, a model pretrained on Pistachio by Schwaller et al. (https://rxn4chemistry.github.io/rxnfp/), for preliminary comparisons. According to our initial analysis presented in Tables S3 to S5, replacing the USPTO database with Pistachio for pretraining could enhance the performance of virtual screening, especially in the EOO_MEW and Solar tasks. It is important to clarify that this observation does not detract from our original findings. Rather, it reinforces our proof-of-concept that leveraging chemical reaction data is advantageous for the virtual screening of organic materials. We advocate for further investigation into the potential of the Pistachio database by researchers who have access to it.

## CONCLUSIONS

In this study, we have successfully demonstrated the concept of pretraining deep learning models on databases not specifically related to organic materials, such as ChEMBL and USPTO, for the virtual screening of organic materials. Leveraging the unique unsupervised pretraining phase of the BERT model, we demonstrate the feasibility of transfer learning across diverse chemical domains, including organic materials, drug-like small molecules, and chemical reactions. Our findings reveal that among the various BERT models pretrained on different databases, those pretrained using molecular SMILES data extracted from the USPTO database exhibited superior predictive performance in the majority of virtual screening tasks. This enhanced performance can be attributed to the USPTO database's broader variety of organic building blocks and its more extensive coverage of chemical space. Our study underscores the potential of applying cross-domain transfer learning to address the challenge of data scarcity in the virtual screening of organic materials and possibly other chemical categories. By showcasing the effectiveness of pretraining deep learning models on diverse chemical databases, we aim to inspire further research in this direction, encouraging the exploration of more extensive databases like Pistachio and fostering advancements in the field of virtual screening.

## SUPPORTING INFORMATION

The supporting information of this article is available in the figshare repository (https://doi.org/10.6084/m9.figshare.24679305).

# ACKNOWLEDGMENTS

This research was supported by the National Natural Science Foundation of China under Grant No. 22108252, the Joint Funds of the Zhejiang Provincial Natural Science Foundation of China under Grant No. LHDMZ23B060001, and Zhejiang Provincial Key R&D Project (No. 2022C01179)

# COMPETING INTERESTS

The authors declare no competing interests.